\documentclass[11pt,twoside]{article}

\usepackage{asp2006_astroph}
\usepackage{epsf}
\usepackage{lscape}
\usepackage{graphicx}

\begin{document}
\title{Star Formation in the Extreme Outer Galaxy: the IMF in a low
 metallicity environment} 

%
%
%

\author{Chikako Yasui\altaffilmark{1}, Naoto Kobayashi\altaffilmark{1},
Alan T. Tokunaga\altaffilmark{2}, Masao Saito\altaffilmark{3}, and
Chihiro Tokoku\altaffilmark{4}}


\altaffiltext{1}{Institute of Astronomy, Univ. of Tokyo, 2-21-1
Osawa, Mitaka, Tokyo 181-0015, Japan}

\altaffiltext{2}{Institute for Astronomy, Univ. of Hawaii, 2680
Woodlawn Drive, Honolulu, HI 96822, USA}

\altaffiltext{3}{ALMA Project, NAOJ, 2-21-1 Osawa, Mitaka, Tokyo
181-8588, Japan}

\altaffiltext{4}{Subaru Telescope, NAOJ, 650 North A`ohoku Place, Hilo,
HI 96720, USA}


\begin{abstract} 

We are conducting a deep near-infrared (NIR) imaging survey of young
embedded clusters in the extreme outer Galaxy (hereafter EOG), at the
Galactic radius ($R_{\rm g}$) of more than 18 kpc. 
The EOG is an excellent laboratory to study the nature of the IMF in a
low-metallicity environment with a great advantage of the proximity
compared to nearby dwarf galaxies, such as LMC \& SMC.
As a first step, we obtained deep NIR images of Digel Cloud 2 clusters
at $R_{\rm g} \simeq 19$ kpc using the Subaru 8.2-m telescope.
The observed $K$-band luminosity function shows that {\it IMF in the low
metallicity environment down to $\sim 0.1$ $M_\odot$ is not
significantly different from the typical IMFs in the field and in the
nearby star clusters} as was suggested in our earlier work
\citep{Yasui2006}.

\end{abstract}

\section{Introduction}


The initial mass function (IMF) in nearby galaxies has been extensively
studied with high-resolution imaging using {\it HST}
\citep[e.g.,][]{{Sirianni2000},{Gouliermis2005}}.
%
It is critical to reach to the substellar mass regime for the complete
census of the IMF, because
{\it a characteristic peak mass, which is important to determine the
shape of the IMF, is at around $\sim$ 0.5 $M_\odot$}
\citep{AnnualReview}. However it is difficult to reach this limit even
for the nearest galaxies with active star formation, such as LMC \& SMC,
because of the lack of spatial resolution.
%
Also, observations of embedded clusters at NIR wavelengths are more
ideal for investigating the ``true'' IMF right after the cluster
formation without being hampered by dust extinction
\citep{AnnualReview}.  Therefore, we started a program to study the IMF
of the young embedded clusters in the EOG defined here as the region at
$R_{\rm g}$ greater than 18 kpc.
The heliocentric distance to EOG is as close as 10 kpc, which is five
times closer than LMC ($D \sim 50$ kpc).
The EOG has a similar environment as nearby irregular dwarf galaxies
 (Im) and damped Ly-$\alpha$ systems (DLAs): low metallicity
 \citep[$\sim - 1$ dex; e.g.,][]{Rudolph2006}, low gas density ($< 0.01$
 cm$^{-3}$), and no or little perturbation from the spiral arms.

\section{Observation}

As a first step for studying the IMF in such environments, we obtained
$JHK_S$ deep images of Cloud 2-N (see Kobayashi et al. in this volume)
using the Subaru Telescope MOIRCS wide-field NIR imager.  Cloud 2 was
identified by Digel et al. (1994) and is located at $R_{\rm g} \sim 19$
kpc and has a metallicity [M/H] = $-0.6$ dex.
With the limiting magnitudes of $J = 23.2$, $H=22.3$, and $K_S = 22.2$
mag, this is the deepest imaging of an EOG cluster, which reaches the
substellar mass regime ($\sim 0.04$ $M_\odot$ for 1 Myr).

\section{IMF}

We identified the cluster members and compared the observed $K$-band
luminosity function (KLF) with the model KLFs (Figure 1) in the same way
as in Yasui et al. (2006, 2008) except that we used the Gaussian IMF
\citep{Miller1979}: $d N / d \log M = A \exp \{-c_1 ( \log M -
c_2)\}^2$ for simple numerical fitting.
Assuming the age of the cluster as $\sim 0.5$ Myr (see Kobayashi et
al. in this volume), the most reliable IMF for the age was identified by
$\chi^2$ minimization.
The best fit

\vspace{0.9mm}
\hspace{-0.75cm}
%
\begin{minipage}[l]{5.2cm}
\begin{flushleft}
KLF and the resultant IMF are shown in Figure 1 (black
lines). Our analysis shows 
that {\it the IMF of the Cloud 2-N cluster down to the substellar mass
regime is consistent with the ``universal IMF'', which is observed in
the solar neighborhood,} confirming our earlier suggestion (Yasui et
al. 2006).
We are continuing the similar study of more EOG embedded clusters to
 understand the dependence of the IMF on various environmental
 parameters.

\end{flushleft}


\end{minipage}
\hspace{3mm}
\begin{minipage}{7.8cm}
\vspace{1mm}
   \begin{center}
    \includegraphics[width=8cm, height=2.9cm]{yasui.eps}
   \end{center}
\vspace{-3mm}
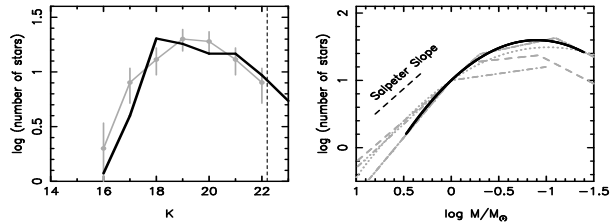
 {\label{fig:bestIMF} {\bf Figure 1.} {Results of KLF fitting assuming
an age of 0.5 Myr for Cloud 2-N.  The best fit KLF and resultant IMF are
shown with black lines in the left and right panels,
while the typical ``universal'' IMFs from literature (see Yasui et
al. 2006 for the detail) are shown with gray lines.}}
\end{minipage}





\end{document}